\documentclass[traditabstract]{aa} 
\usepackage{graphicx}
\usepackage{txfonts}
\begin{document}
   \title{The HARPS search for southern extrasolar planets
   \thanks{Based on observations made with the HARPS instrument 
	 on the ESO 3.6 m telescope at La Silla (Chile), under the GTO program ID
	    072.C-0488 and the regular programs: 085.C-0019, 087.C-0831 and 089.C-0732. }
	    }
   \subtitle{XXXIII. New multi-planet systems in the HARPS volume limited sample:\\
   	a super-Earth and a Neptune in the habitable zone.
   	\thanks{RV data are only available in electronic form at the CDS}
   }

   \author{
   G.~Lo~Curto\inst{1}
\and M.~Mayor\inst{2}
\and W.~Benz\inst{3}
\and F.~Bouchy\inst{2,5}
\and G.~H\'ebrard\inst{5,6}
\and C.~Lovis\inst{2}
\and C.~Moutou\inst{4}
\and D.~Naef\inst{2}
\and F.~Pepe\inst{2}
\and D.~Queloz\inst{2}
\and N.C.~Santos\inst{7,8}
\and D.~Segransan\inst{2}
\and S.~Udry\inst{2}
           }

\offprints{G.~Lo~Curto\\e-mail: {\tt glocurto@eso.org}}

\institute{
ESO - European Southern Observatory, Karl-Schwarzschild-Strasse 2, 85748 Garching bei M\"unchen, Germany
\and
Observatoire de Gen\`eve, Universit\'e de Gen\`eve, 51 ch. des Maillettes, 1290 Sauverny, 
Switzerland
\and
Physikalisches Institut Universit\"at Bern, Sidlerstrasse 5, 3012
Bern, Switzerland
\and
Laboratoire d'Astrophysique de Marseille, Traverse du Siphon, 13376
Marseille 12, France
\and
Institut d'Astrophysique de Paris, UMR7095 CNRS, Universit\'e Pierre \& Marie Curie, 
98bis boulevard Arago, 75014 Paris, France 
\and
Observatoire de Haute-Provence, CNRS/OAMP, 04870 Saint-Michel-l'Observatoire, France
\and
Centro de Astrofisica, Universidade do Porto, Rua das Estrelas, 
4150-762 Porto, Portugal
\and
Departamento de F{\'\i}sica e Astronomia, Faculdade de Ci\^encias, Universidade do Porto, Portugal
}

   \date{}
   


    \abstract
{
The vast diversity of planetary systems detected to date is defying our capability of understanding their formation and evolution.
Well-defined volume-limited surveys are the best tool at our disposal to tackle the problem, via the acquisition of robust statistics 
of the orbital elements.
We are using the HARPS spectrograph to conduct our survey of $\approx 850$ nearby solar-type stars, 
and in the course of the past nine years we have monitored the radial velocity of HD103774, HD109271, and BD-061339. 
In this work we present the detection of five planets orbiting these stars, with $m \sin(i)$ between 0.6 and 7 Neptune masses,
four of which are in two multiple systems, comprising one super-Earth and one planet within the habitable zone of a late-type dwarf.
Although for strategic reasons we chose efficiency over precision in this survey,
we have the capability to detect 
planets down to the Neptune and super-Earth mass range,  as well as multiple systems, provided that enough data 
points are made available.

}


   \maketitle
   
%

\section{Introduction}

After more than 15 years of exo-planets detections (\cite{peg51}, \cite{schneider}), 
we have witnessed an extreme diversity of planetary systems, 
but we have not yet detected  a system that could be considered a twin to our own solar system, not even limited
to the giant (and easier to detect) planets. We are left with the question of whether our system, and with it our planet, 
are the norm, or rather the exception in planet formation and evolution.
To date, only about one third of the detected planets are found in confirmed multi-planet systems (\cite{schneider}). This fraction is steadily 
rising with the increasing amount of available high-quality data, indicating that we are fighting against a detection bias.
Multi-planet systems are more challenging to detect because they require many data points to permit the identification 
of the individual signals and because, as is revealed from the data collected so far, often they involve low-mass 
planets in the super-Earth to Neptune mass range (e.g. \cite{locurto2010}, \cite{hebrard2010},  \cite{lovis11}, \cite{fischer12}). 
The combination of the low amplitude of these signals with the more complex motion induced on the parent star by the systems requires a substantial observational effort both in terms of observing time and of the measurement precision.

Our survey is using the HARPS spectrograph, which is mounted full time at the 3.6m ESO telescope in La Silla and is 
optimized for precise radial velocity measurements (\cite{mayor03}).
To date, it has discovered more than 150 exo-planets, most of them (through the high-precision program) 
with masses in the super-Earth range; this definitely changed our view on the field.

Since the start of HARPS operations we are monitoring about 850 stars with moderate radial velocity precision ($\approx 2-3m s^{-1}$), searching for Jupiter and Neptune planets orbiting solar-type stars (\cite{locurto2010}), while the HARPS high-precision program (\cite{mayor03}), which routinely achieves precision
better than $1ms^{-1}$, targets lower mass planets, mostly in the super-Earth mass regime. The goals of our program are the acquisition of accurate orbital elements 
of Jupiter and Neptune-mass planets in the solar neighborhood (within 57.5pc) and the search for Jupiter twins, which 
might reveal solar system twins. Of the 43 planets discovered so far by this program, 17 have periods longer than 1000 days (\cite{moutou09}, \cite{naef10}, two more are presented  in a paper in preparation), 
but their eccentricities are significantly higher than that of Jupiter. The program has detected three systems with two planets and one triple planetary system so far (\cite{hebrard2010}, \cite{locurto2010}).
Out of the nine planets in multiple systems we have discovered so far, only two have masses larger than or equal to Jupiter, 
which is consistent with the observed trend of low-mass planets being detected primarily in multiple systems.

In this paper we report three stars with multiple Keplerian-like signals. 
In section 2 we discuss the stars' properties, and in section 3 the orbital solutions.
In section 4 we briefly discuss the encountered systems.


\section{The stars}

The three stars belong to our volume-limited sample and were selected to be low-activity, non-rotating, solar-type stars on the main sequence (\cite{locurto2010}).
We have used the magnitude estimates by Hog (2000), and the one by Koen (2011) for the redder star BD-061339. 
They originate from the  Tycho and HIPPARCOS catalogs, respectively (\cite{hip97}, \cite{hip07}).
The color indices and the parallaxes are from \cite{hip07}, and the bolometric corrections are extracted from a calibration to the data presented 
by Flower (1996), with uncertainties generally below 0.1 mag.

\subsection{HD103774}




The star HD103774 is a young, metal-rich, main-sequence star with spectral type F5V (\cite{pickles10}, \cite{sousa11}). Its average activity level is low, but it varied considerably during 
the time span of our observations (see figure \ref{hd103774rhk}), going from a more "active" state with $log(R_{HK})=-4.72$ to a more "quiet"
state with $log(R_{HK})=-5.00$ (see \cite{noyes84} for a definition of the activity indicator). We therefore expect to measure some level of  radial velocity jitter.

\begin{figure}[h]
   \centering
   \includegraphics[width=8.5cm, height=5.3cm]{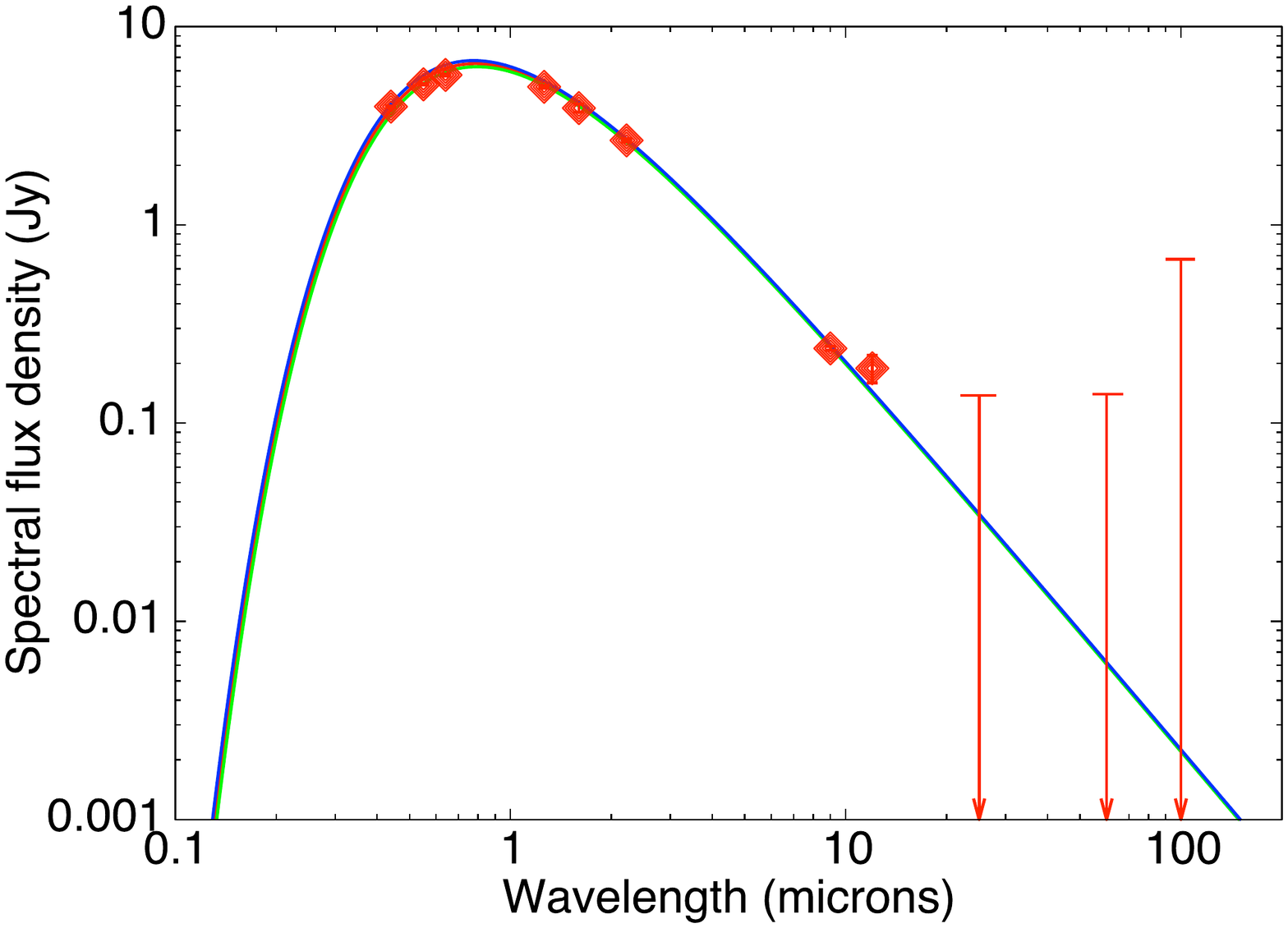}
      \caption{Spectral flux density for HD103774 measured in Jy. The solid line is a black body with the temperature 
                      of the star: 6489K.  The arrows indicate that the three data points are only upper limits. }
         \label{hd103774flux}
  \end{figure}

Using the online data service VizieR, we found measurements of this star in the mid- and far-infrared, from the IRC (instrument onboard the AKARI satellite,  \cite{irc2010}) 
and IRAS (\cite{moshir90}) catalogs (see figure \ref{hd103774flux}). The data points at 25, 60, and 100 $\mu m$ are upper limits. The solid line in figure \ref{hd103774flux} 
reproduces the black body corresponding to the star's photospheric temperature of $6489 K$. The uncertainty on the temperature (77K) would not be noticed on the scale of this graph. 
The point at $12\mu m$ is deviating from the black body and might seem to indicate an infrared excess. Circum-stellar dust is indeed quite common around early-type young stars.
While the evidence for infrared excess at 12 $\mu m$ is only marginal (1 sigma) and calls for more precise measurements, no excess is measured up to 
9 $\mu m$, giving an indication on the minimum size of the grains that could possibly be present in the disk.


\subsection{HD109271}


This old and quiet G5V star does not show infrared excess (\cite{irc2010}, IRAS data are not available),  and has a near-solar metallicity (\cite{sousa11}). Its basic parameters, as for the other stars, are shown in table \ref{TableStar}.

\subsection{BD-061339}



BD-061339 is a late-type dwarf (spectral type K7V/M0V), with no obvious infrared excess (\cite{moshir90}, IRAS Faint Source Reject Catalog, Version 2.0).

The estimation of the metallicity in late-type dwarfs is  complicated by the abundance of molecular bands of various metal oxides in the spectra  (e.g. TiO, VO).
Neves et al. (2012) have discussed various calibrations present in the literature and presented an improved photometric calibration for the metallicity.
We adopted their prescription to estimate the metallicity of BD-061339, and used the dispersion around their calibration as an estimate of the uncertainty.
This value is well above the measurement uncertainty computed via error propagation from their formula, and is limited by the quality of the calibration.

Owing to the red color of the star the estimate of the activity index $\log R'_{\mathrm{HK}}$ is considered unreliable and no value is included in  table \ref{TableStar}.


\begin{table*}
\caption{Observed and inferred stellar parameters of the three stars of this study. $\pi$: parallax, $d$: distance, $M_V$: absolute magnitude in V, $Bol.$ $corr.$: bolometric correction, 
$L$: luminosity, $T_{eff}$: effective stellar temperature, $M_*$: stellar mass, log $g$ : gravity at the star's surface, v$\sin{i}$: projected rotational velocity of the star, 
$\log R'_{\mathrm{HK}}$ : activity indicator (\cite{noyes84}), $P_{\mathrm{rot}}$($\log R'_{\mathrm{HK}}$): star's rotation period estimated via the activity indicator.
}
\label{TableStar}
\centering
\begin{tabular}{l l r r r r r r l}
\hline\hline
\multicolumn{2}{l} {\bf Parameter} &\hspace*{2mm} & \bf HD\,103774  &\hspace*{2mm} & \bf HD\,109271  &\hspace*{2mm} & \bf BD-061339 & \hspace*{2mm} \\
\hline

Hipparcos name 	&	HIP		& & 58263			& & 61300			& & 27803 			& \\
Spectral type 		& 			& & F5V 				& & G5V 				& & K7V / M0V 			& \\
$V$ 				& [mag] 		& &  7.12 $\pm$ 0.01	& & 8.05 $\pm$ 0.01 	& & 9.69 $\pm$ 0.01 	&  (\cite{hog2000},  \cite{koen2011}) \\
$B-V$			& [mag] 		& & 0.503 $\pm$ 0.014 	& & 0.658$\pm$0.002 	& & 1.321 $\pm$ 0.001	&  (\cite{hip07}) \\
$\pi$ 			& [mas] 		& &18.03$\pm$0.52		& & 16.21$\pm$0.75 	& & 49.23$\pm$1.65 	& (\cite{hip07}) \\
d 				& [pc] 		& &  55$\pm$2 			& & 62$\pm$3 			& & 20$\pm$1  			& \\
$M_V$ 			& [mag] 		& & 3.40 $\pm$ 0.06 	& & 4.1 $\pm$ 0.1 		& & 8.15 $\pm$ 0.07 	& \\
Bol. Corr. 			& [mag]		& & -0.013			& & -0.097			& & -0.782 			& (\cite{flower96}) \\
$L$ 				& [$L_{\odot}$] 	& & 3.5 $\pm$ 	0.3		& & 2.0 $\pm$ 	0.3		& & 0.095 $\pm$ 0.01 	& \\
$T_{\mathrm{eff}}$ 	& [K] 			& & 6489 $\pm$ 77 		& & 5783 $\pm$ 62 		& & 4324 $\pm$ 100 	& (\cite{sousa11}, \cite{tyco2_06}) 	 \\
$\mathrm{[Fe/H]}$	& [dex]	 	& & 0.28 $\pm$ 0.06		& & 0.10 $\pm$ 0.05 	& &-0.14$\pm$0.17		&  (\cite{sousa11}, \cite{neves12})  \\
$M_*$ 			& [$M_{\odot}$] & &1.335$\pm$0.03		& & 1.047$\pm$ 0.024 	& & 0.7 				& (\cite{girardi2000}) \\
log $g$  			& [cgs] 		& & 4.26 $\pm$0.13		& & 4.28 $\pm$ 0.10	 	& & 4.74 $\pm$ 0.02		& (\cite{girardi2000}, \cite{sousa11})  \\
age  				& [Gyr] 		& & 1.05 $\pm$ 0.64 	& & 7.3 $\pm$ 1.2 		& & 4.4 $\pm$ 4.0		& (\cite{girardi2000})  \\
v$\sin{i}$ 			& [km s$^{-1}$] 	& & 8.1 				& & 2.7 				& & 0.57 				& (\cite{gleb05}, CORAVEL) \\
$\log R'_{\mathrm{HK}}$ 	& 		& & -4.85 $\pm$ 0.07 	& & 	-4.99 $\pm$ 0.08	 & & -  				& \\
$P_{\mathrm{rot}}$($\log R'_{\mathrm{HK}}$) 
				& [days] 		& & 7				& & 24 				& & -					& (\cite{noyes84}) \\
\hline
\end{tabular}
\end{table*}


\section{Orbital solutions}

We studied the radial velocity (RV) variation of our target stars via the analysis of the periodograms of the RV data 
following the generalized Lomb-Scargle periodogram method outlined by Zechmeister and Kuerster (2009) and via the use of genetic algorithm as described in the work by Segransan et al. (2011). In all cases we verified whether the RV was correlated with the bisector span of the cross correlation function (CCF), the activity indicator $R_{HK}$ (\cite{noyes84}), and the FHWM of the CCF before claiming a companion. In addition to this, we also studied the periodogram of these indicators.
When attempting a fit with more degrees of freedom (e.g., using a model with more planets or adding a long-term drift to the model), 
we verified with an F-test that the new fit brings a significant improvement over the fit with fewer degrees of freedom.
We ran an F-test and obtained a false-alarm probability (FAP) for which we used a threshold of $10^{-4}$ to identify significant peaks.
The orbital elements are finally displayed in table \ref{orbittab}.

\subsection{HD103774}

\begin{figure}
   \centering
   \includegraphics[width=9cm]{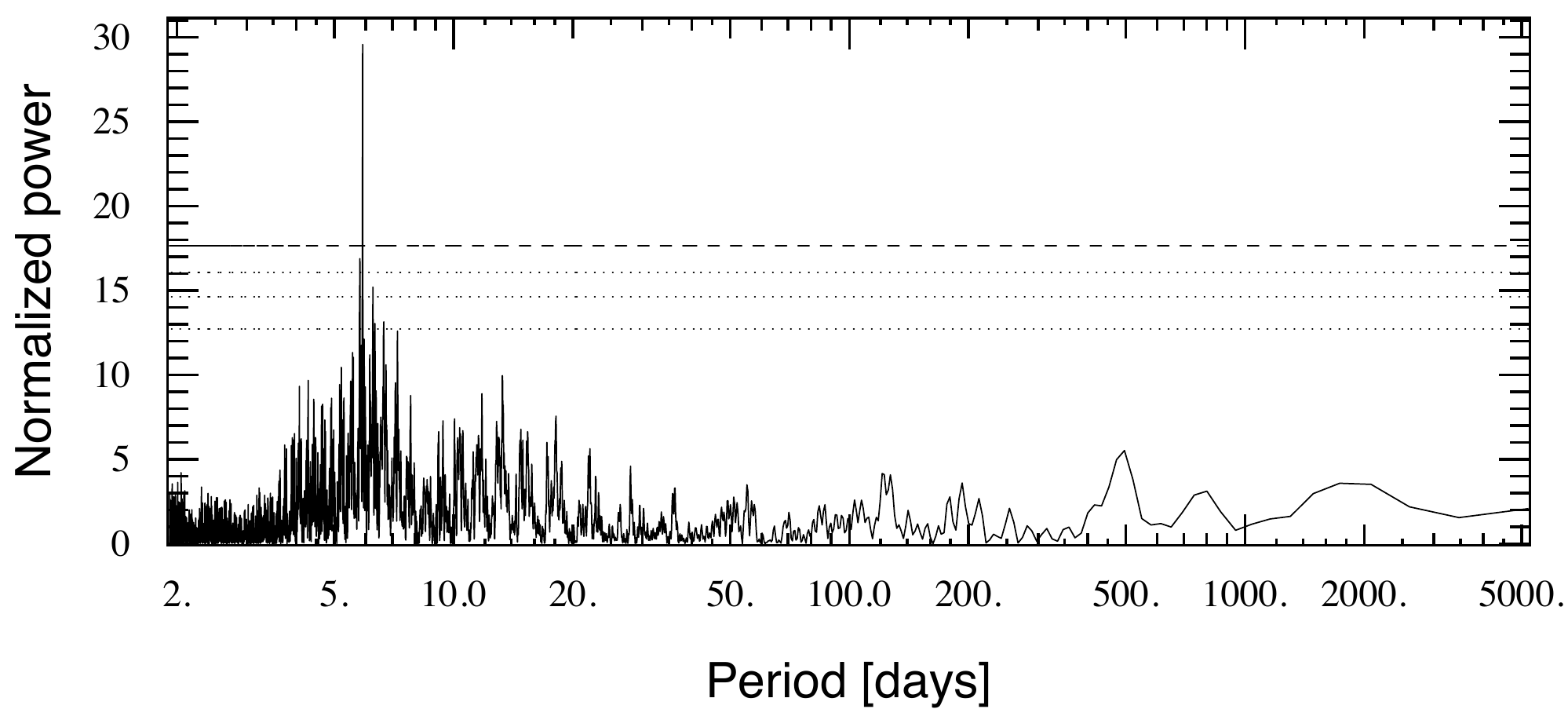}
      \caption{Periodogram of the RV data for HD103774. A significant peak 
	is visible at $\approx 5.9$ days. The topmost dashed line is the $10^{-4}$ false-alarm probability, 
	the lower dotted lines correspond to $10^{-3}, 10^{-2}$, and $10^{-1}$ false-alarm probabilities.             }
         \label{hd103774per}
  \end{figure}

We acquired 103 radial velocity measurements for this star in about 7.5 years.
The average uncertainty is $\approx 2.6m s^{-1}$, while the standard deviation of the radial velocity over the entire data set is $\approx 26 m s^{-1}$, a clear indication of variability. 
The periodogram of the radial velocities (Fig. \ref{hd103774per}) indicates an excess of power at $\approx 5.9$ days, 
while no other significant peak is visible. 
The one-year and the two-year aliases (at 5.79d and 5.84d) are also present in the periodogram, but at half the power of the main signal.

\begin{figure}
   \centering
   \includegraphics[width=8cm]{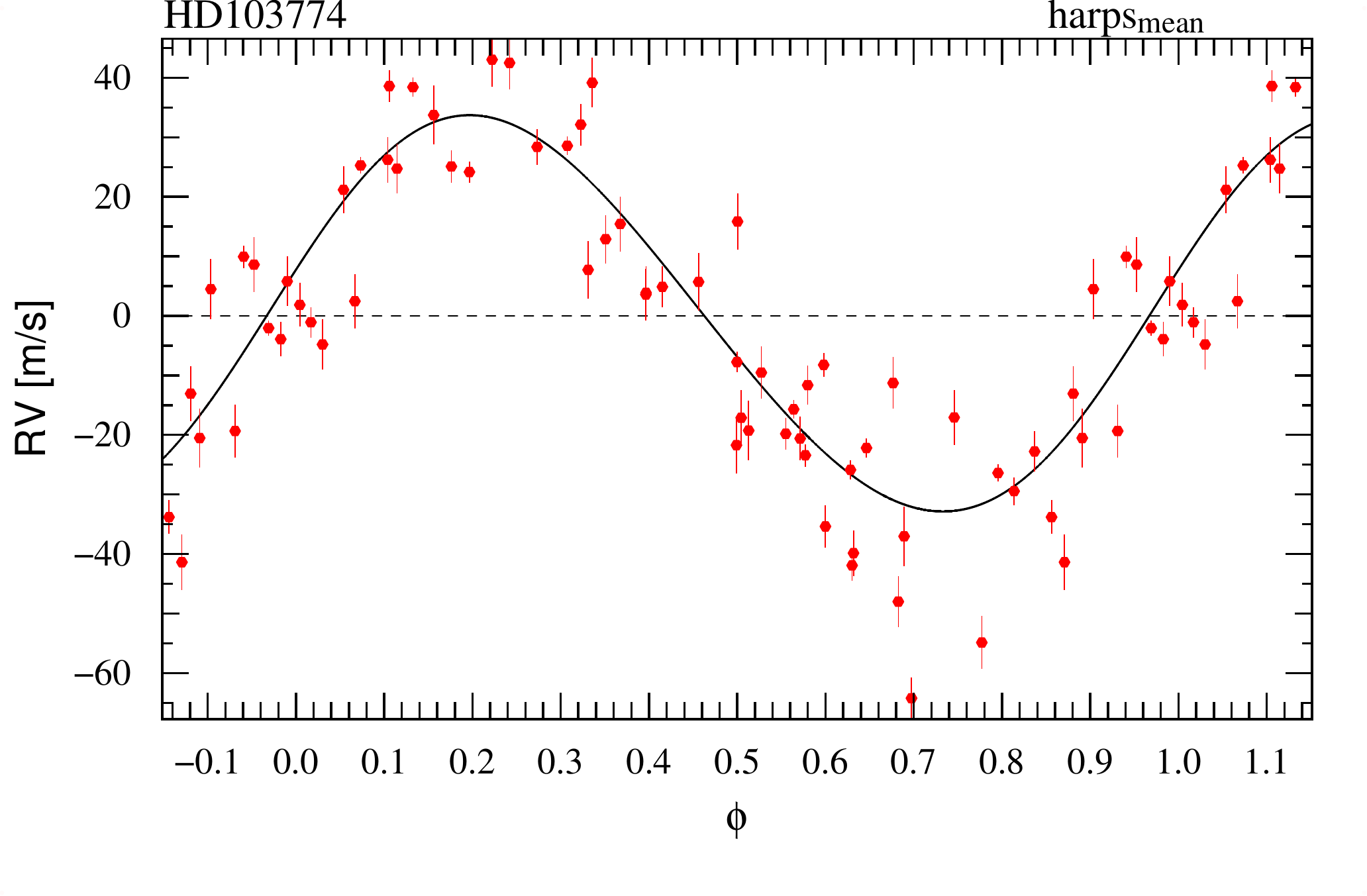}
      \caption{Keplerian fit of the radial velocity data of HD103774 as a function of orbital phase.  }
         \label{hd103774pha}
  \end{figure}

In figure \ref{hd103774pha}  we show the Keplerian fit as a function of phase together with the 
data. The orbit is evident, although noisy.

\begin{figure}
   \centering
   \includegraphics[width=8cm]{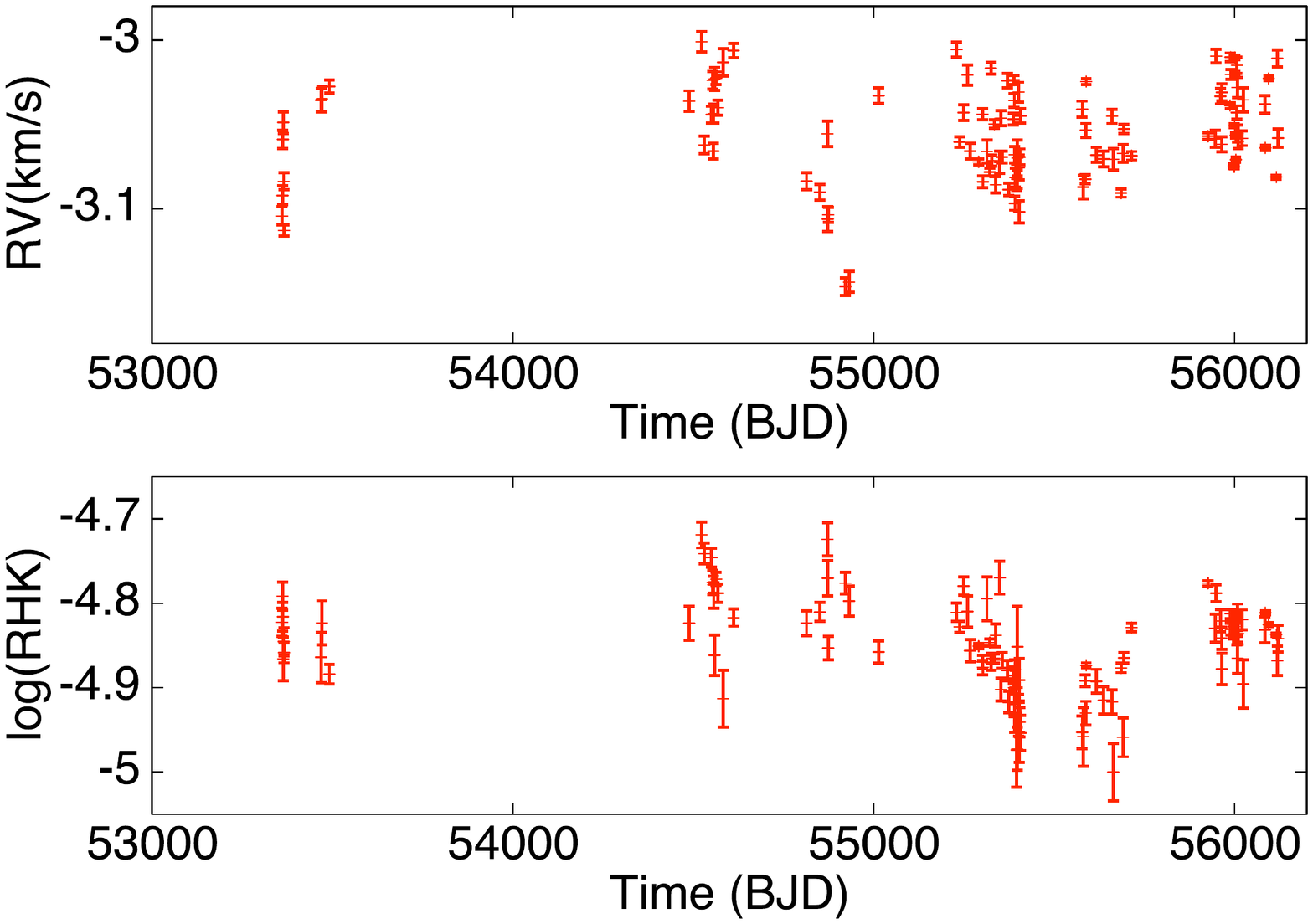}
      \caption{Variation of the radial velocity (upper panel) and the activity index (lower panel) of HD103774 with time.  }
         \label{hd103774rhk}
  \end{figure}

Although the periodogram shows various structures ("bumps" at $\approx$ 11d, 13d, 15d, 470d, 740d), they are not significant, and 
we see no correlation of the radial velocities with the activity indicators $S_{MW}$ (\cite{baliunas95}) or $log(R_{HK})$ (Pearson coefficient $<0.1$),
nor with the bisectors or the FWHM. The time series of the activity index $log(R_{HK})$ shows strong variations
of the index with time (Fig. \ref{hd103774rhk}), indicating that the star activity level might increase considerably.

Because we have enough data points, we separately analyzed three groups of RV data  
in three disjoint Julian date bins. From orbital fitting we have obtained the same periods
for the three groups of data within the experimental uncertainty of $\approx 10^{-3}$ days. The same is true for 
the phase of the orbit, although in this case the uncertainties are larger ($\approx 2$ days). Therefore the
detected period is constant within the time span of the measurement, i.e., 7.5 years.

The period of $\approx 5.9$ days is quite close to the estimated rotation period of the star (\cite{noyes84}) of  $\approx 7$ days.
However, the lack of periodicity of the bisector span, together with the lack of correlation between bisector and
radial velocity, and the fact that the orbital period is strictly constant along the 7.5 years of data gives us confidence that
the most likely cause of the radial velocity variation with the period of $\approx 5.9$ days is a low-mass companion.

\begin{figure}
  \centering
  \includegraphics[width=8cm]{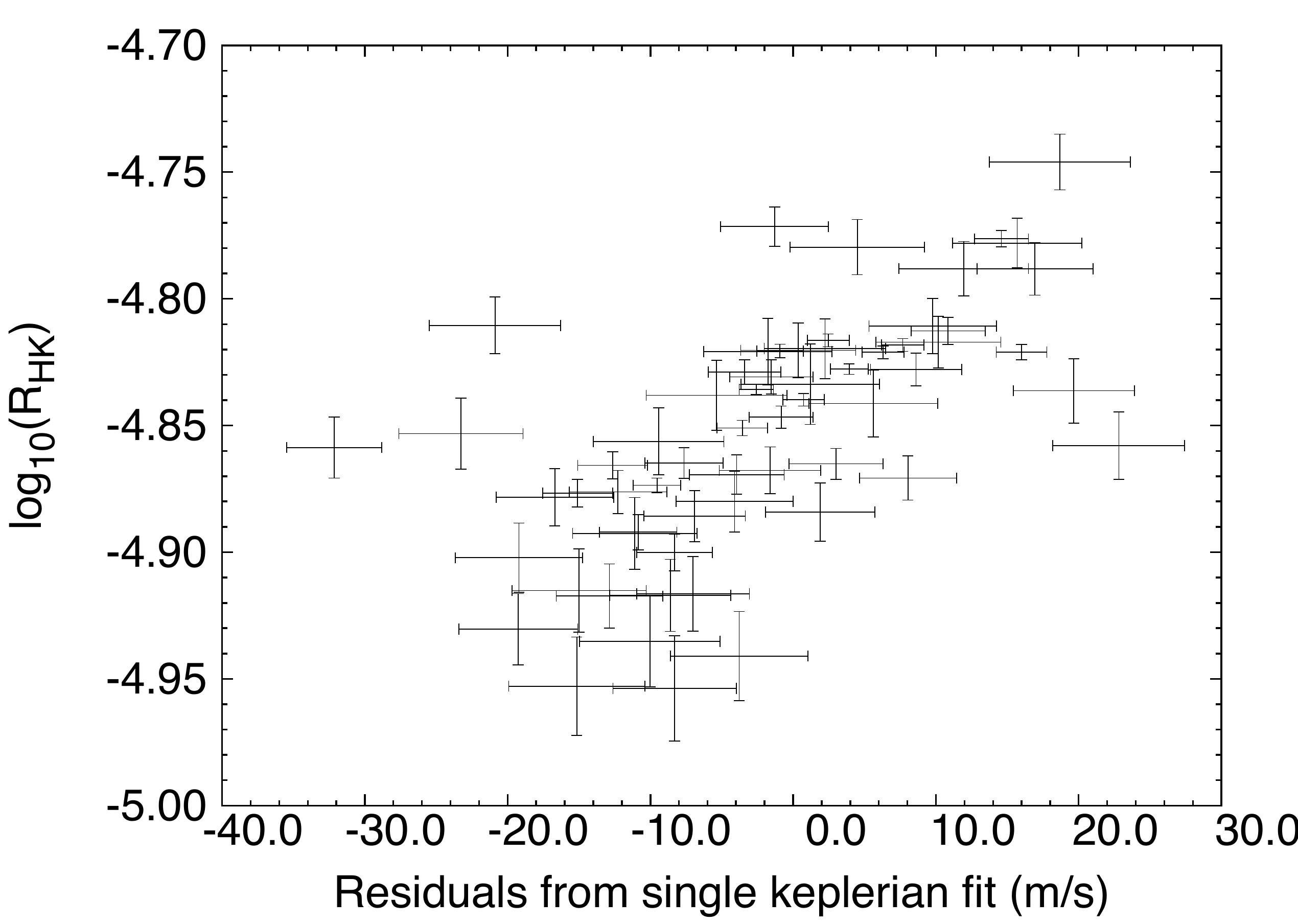}
     \caption{Activity index $log(R_{HK})$ of HD103774 as a function of the residuals from the single-Keplerian fit.
             A strong correlation is evident. The value of the Pearson correlation coefficient is 0.6.}
        \label{hd103774omcrhk}
 \end{figure}

When looking at the residuals we still notice a significant scatter of the data.
Moreover, the periodogram of the residuals shows an excess of power corresponding to a period of 
$\approx 750$ days, jointly with its one-year alias at $\approx 250$ days.
We significantly improved the overall fit to the data by fitting two Keplerians.
However, we detected a strong correlation between the radial velocity residuals to the single-Keplerian fit
and the activity index $log(R_{HK})$ (Fig. \ref{hd103774omcrhk}) 
with the value of the Pearson correlation coefficient which is equal to 0.6,
and found a peak in the periodogram of the activity index corresponding to this period.
For these reasons we attribute the variation of the residuals to the one-Keplerian fit to stellar magnetic cycles (\cite{santos10}).

The highly varying $log(R_{HK})$ reaches values as high as  $-4.7$, 
lending further support to the suggestion
that the activity of this early-type, relatively young star could be responsible for the large 
$\chi^2$ and the scatter of the residuals around the one-planet fit.

\subsection{HD109271}

\begin{figure}
  \centering
  \includegraphics[width=8cm]{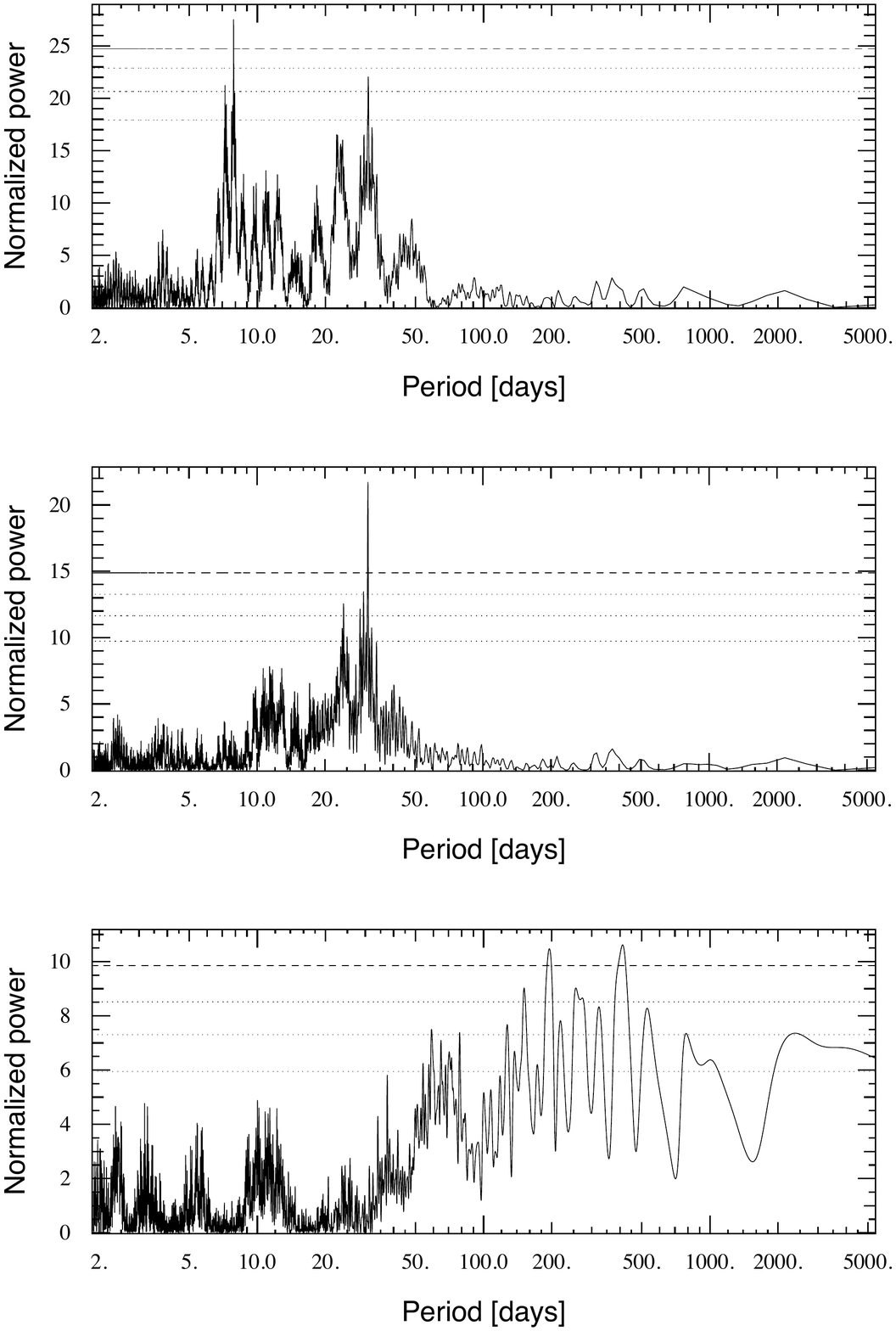}
     \caption{Periodograms of HD109271. From top to bottom: periodogram of the data,
                     the residuals after the one-planet fit, and the residuals after the two-planets fit.
                     The dashed line marks the $10^{-4}$ FAP, the dotted lines, 
                     from top to bottom, the $10^{-3}, 10^{-2}$, and $10^{-1}$ FAP. }
        \label{hd109271per}
 \end{figure}

We measured the radial velocity of the star HD109271 95 times in 7.5 years.
The RV varies by $6ms^{-1}$ RMS, well in excess of the average RV error, which is below $1ms^{-1}$.
The periodogram of the observed data is shown in the top panel of figure \ref{hd109271per}, and two peaks are clearly visible, 
one of which is above the $10^{-4}$ FAP level. We fit a Keplerian to this signal and obtained a good 
fit for a period of $\approx 7.85$ days. 


After fitting the first period, looking at the periodogram of the residuals (figure \ref{hd109271per}, middle panel) we notice a second peak, 
corresponding to the second-highest peak in the periodogram of the observations, above our FAP limit.
This signal corresponds to a variation with a period of $\approx 30.98$ days.


\begin{figure}
  \centering
  \includegraphics[width=8cm, height=8cm]{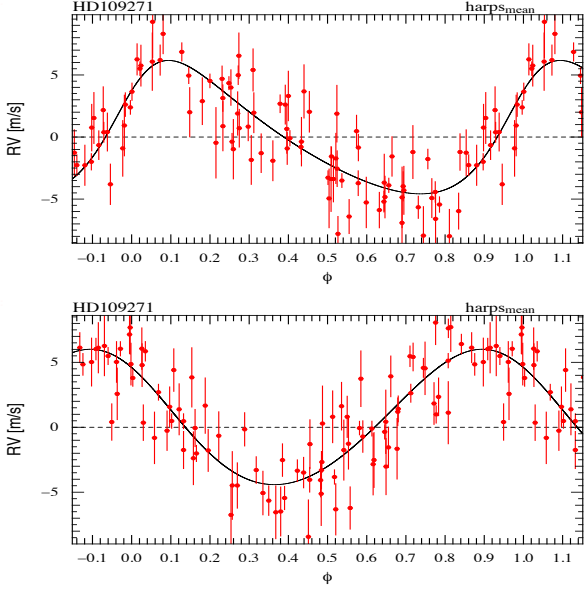}
     \caption{Radial velocity curves of planets HD109271b (top panel) and HD109271c
     (bottom panel) as a function of phase.}
          \label{hd109271_pls}
 \end{figure}

The phased radial velocity curves of the two planets resulting from a two-Keplerians (simultaneous) fit and after
subtraction of the RV signal of the other planet are shown in figure \ref{hd109271_pls}.
Neither of the signals is correlated with the bisector, with the activity index, or the FWHM of the CCF;
we also verified that the periodograms of these parameters have no power at periods corresponding 
to the identified orbital periods. 
Moreover, both signals were identified using the entire data set and also 
using three disjoint subgroups of the entire data set, indicating that they are present during the whole time span of our observations.
Therefore we consider the origin of the two signals to be two low-mass companions orbiting the star HD109271.

After the two-planets fit to the HD109271 RV data, the periodogram of the residuals shows two signals above our FAP threshold 
at P$\approx$ 198d and P$\approx$ 430d  (figure \ref{hd109271per}, bottom panel). 
Within the uncertainties, these periods are the one year alias of each other.
When constraining the analysis to consider time spans shorter than one year but longer than 200 days, 
both peaks disappear, and they reappear together only when the time span is longer than $\approx 600$ days,
an indication that the signal could be the peak at $\approx 430$ days, and the one at $\approx 198$ days could be the alias.
The third signal, which could be associated to a 1.3 Neptune-mass planet on a 1 AU eccentric (e$\approx 0.36$) orbit, is very noisy 
and at the limit of detectability by our survey ($K1<2ms^{-1}, <\delta RV>\approx 1ms^{-1}$). Although we cannot associate this signal
to stellar activity, we need a larger data set and a higher precision to eventually confirm it as a companion to HD109271.





\subsection{BD-061339}

\begin{figure}
  \centering
  \includegraphics[width=8cm]{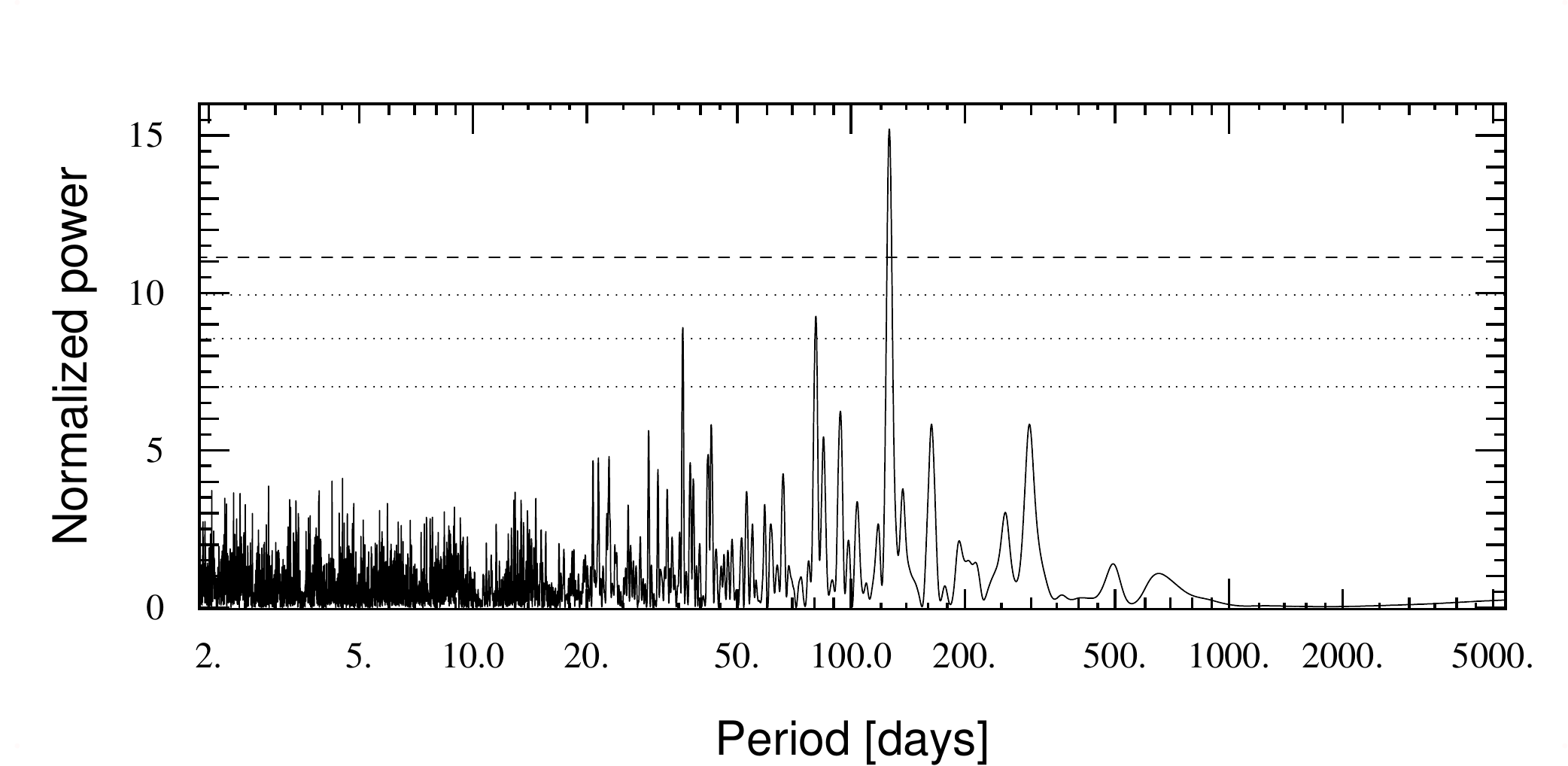}
     \caption{Periodogram of the observations of BD-061339. The period at $\approx$ 125 days is clearly visible well above the $10^{-4}$ FAP level.}
     \label{bd061339_per}
 \end{figure}

We collected 102 radial velocity measurements of the star BD-061339 in about eight years .
The average RV uncertainty is 2.7 ms$^{-1}$. Figure \ref{bd061339_per} shows the
periodogram of the observations, clearly indicating an excess of power at $\approx$ 125 days,
well above the $10^{-4}$ FAP level.

\begin{figure}
  \centering
  \includegraphics[width=8cm]{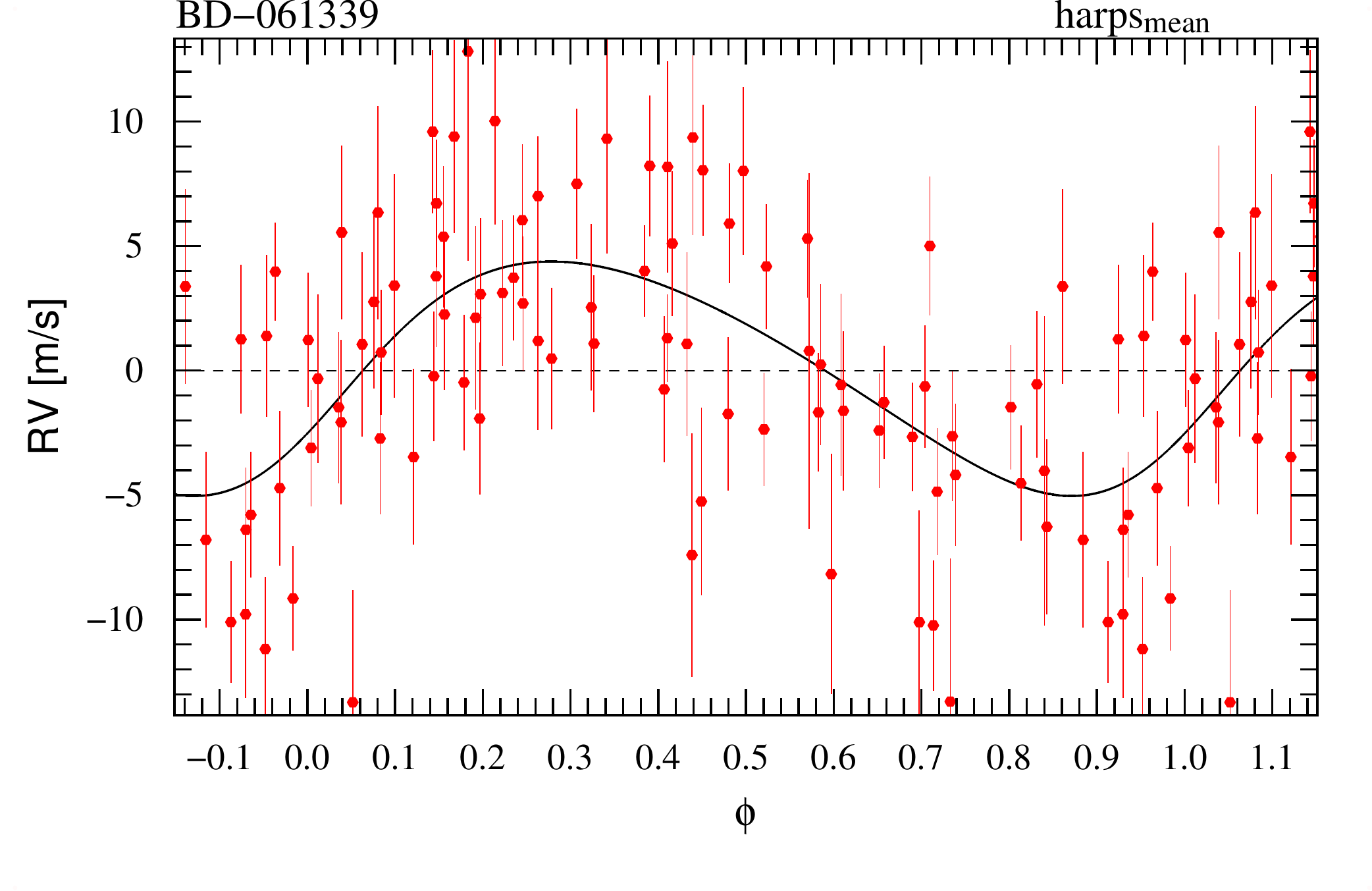}
     \caption{Orbital fit of the 125-day signal to the BD-061339 RV data.}
     \label{bd061339_pl1}
 \end{figure}

A Keplerian fit to the data allowed us to reconstruct an orbit whose RV variation as function of the orbital phase is shown in Figure \ref{bd061339_pl1}.
The orbital fit is very robust, no correlation is seen between the radial velocities and the activity index, the bisector span or the FWHM of the CCF,
and subsamples of the data selected in Julian date show a periodic RV variation with the same period within the uncertainties.

The periodogram of the residuals shows an excess at P$\approx 3.9$ days with the
signal just above the $10^{-4}$ FAP level.

 \begin{figure}
  \centering
  \includegraphics[width=8cm]{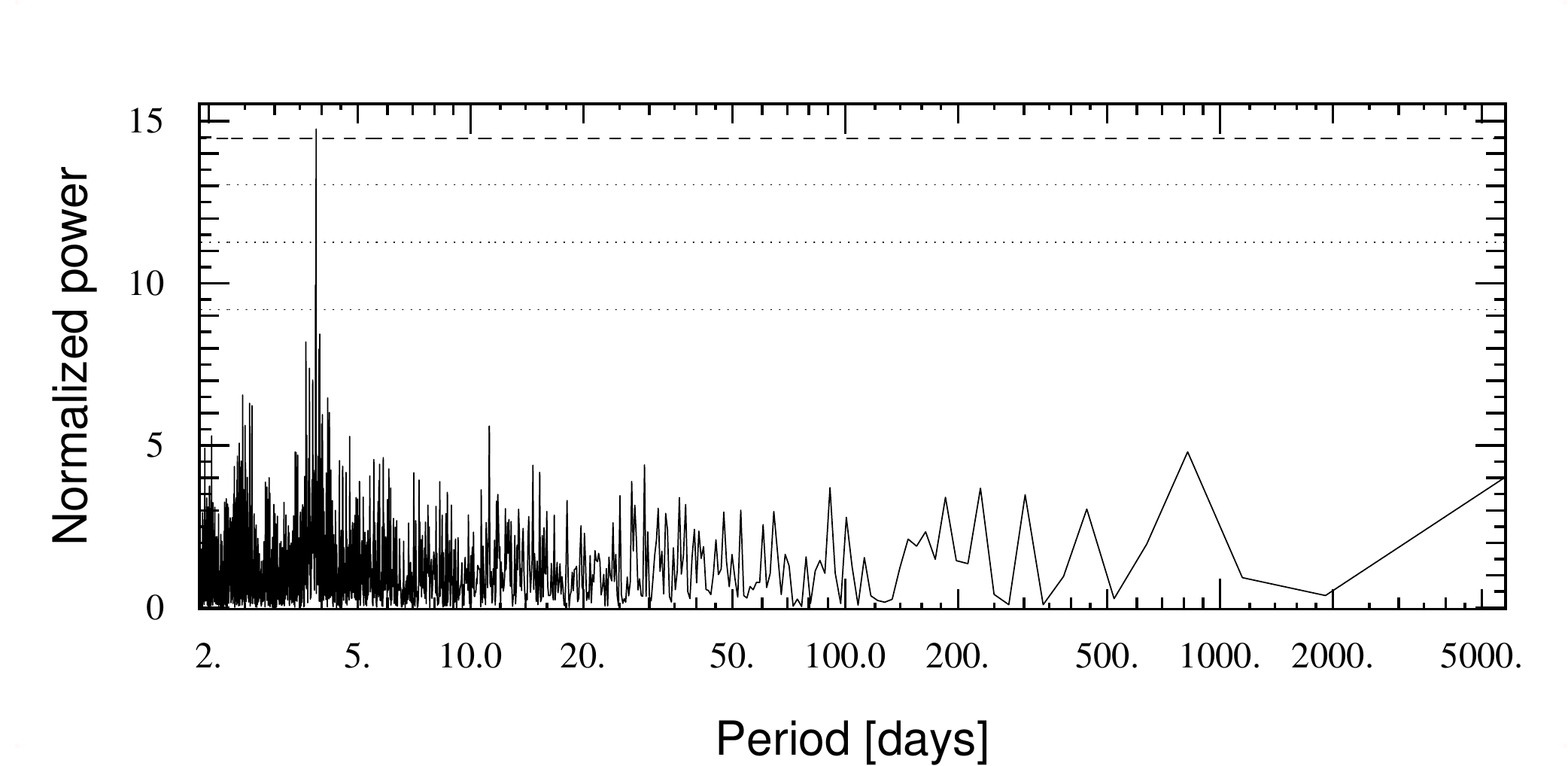}
     \caption{Periodogram of the residuals to the first Keplerian fit to the of BD-061339 RV data. A signal at P$\approx 3.9$ days is visible just above the $10^{-4}$ FAP level.}
     \label{bd061339_per_omc}
 \end{figure}

\begin{figure}
     \centering
     \includegraphics[width=8cm]{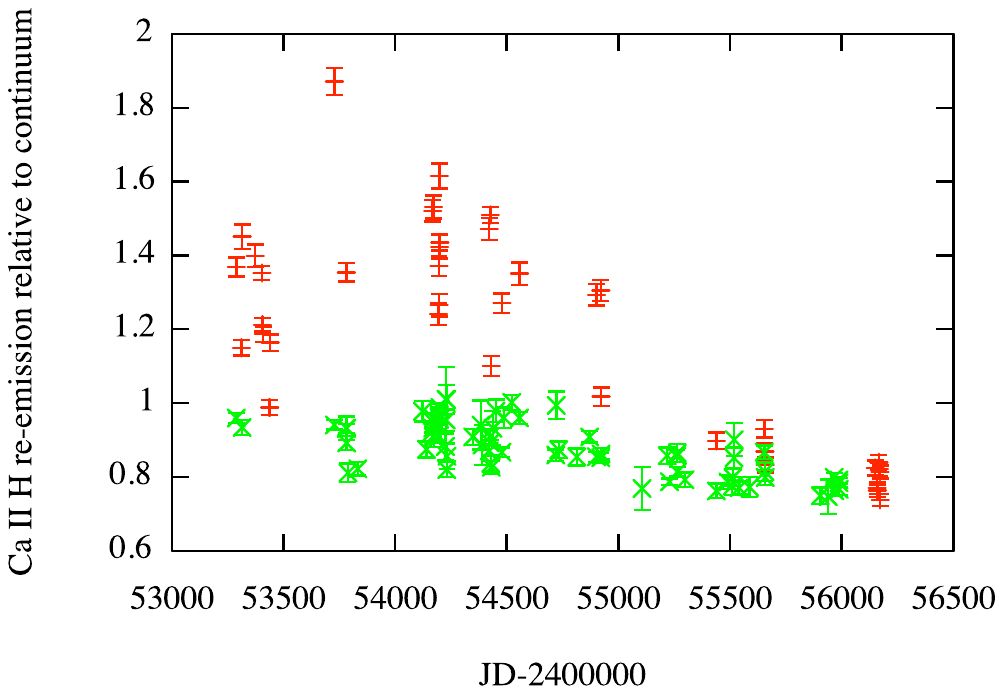}
     \caption{Evolution in time of the integrated flux in the Ca II H line of BD-061339. The star appears much less active during the last observation campaigns.
     The red points (darker) represent the amplitude of the Ca II H re-emission normalized at $6540\AA$, while the green points (lighter) represent the $S_{MW}$ index.
     The latter one is heavily affected by the low S/N in the blue side of the spectrum.}
     \label{bd061339_caiih_t}
 \end{figure}

The peak in the periodogram at $\approx 3.9$ days is well defined, 
and the fit with a two-planets model converges and significantly improves the $\chi^2$.
The F-test gives a $99\%$ probability of improvement of a two-Keplerian simultaneous fit over a one-Keplerian.
The uncertainty on the period from the fit is less than $10^{-3}$ days due to the many
periods covered through the entire observations campaign. 
 Because of the quite red color of the star 
($B-V =1.321$), the activity indicator $log(R_{HK})$ is not reliable.
In an attempt to study the activity of this evolved star, we observed the evolution
of the re-emission in the Ca II H and K lines with time. To quantify the re-emission
we normalized the spectra to the continuum. As suggested by Mould (1976),
the region around $6540\AA$ is only weakly affected by TiO bands. 
By direct inspection we verified that the contribution of the TiO bands at this wavelength for this star is indeed negligible,
and that the width of the $H_{\alpha}$ line is such that it does not affect  the region we planned to use to estimate the continuum.
The normalized integrated flux of the Ca II H line as function of time is shown in figure \ref{bd061339_caiih_t}.
It drops by a factor $\approx 2$ between JD 2454000 and JD 2456000. In figure \ref{bd061339_caiih_t} we also show the 
time dependence of the $S_{MW}$ index: this index drops as well as a function of time, although the slope  is significantly smaller. 
The $S_{MW}$ index is normalized to the continuum much closer to the re-emission line, but the signal-to-noise ratio (S/N) is very low in this region,
and the substantial photon noise strongly influences the measurement. For this reason we attempted the normalization at $6540\AA$.

\begin{figure}
  \centering
  \includegraphics[width=9cm]{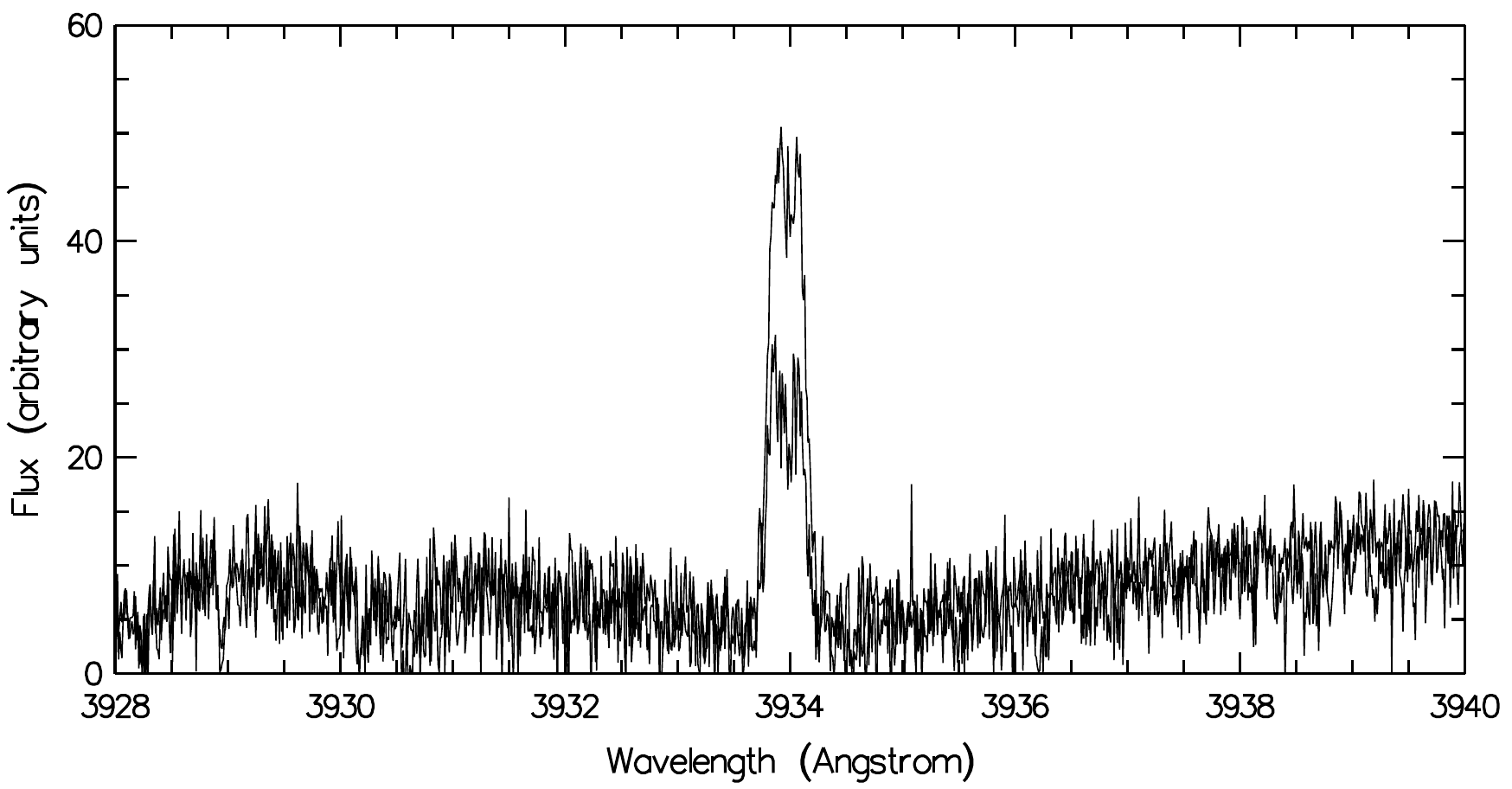}
     \caption{Ca II H re-emission from two spectra of BD-061339 acquired in April 2007 and September 2012.
     The intensity of the line of the more recent spectrum is reduced almost by half.}
     \label{bd061339_caiih}
 \end{figure}

The star  is becoming less active in recent times. Figure \ref{bd061339_caiih} shows the 
re-emission in the Ca II H line in April 2007 (JD = 2454200) and September 2012 (JD = 2456200). 
The intensity of the line almost halved during this period.

Because of the variation of the activity level of the star with time, we divided our data set into two subsamples:
one containing data acquired before JD = 2454500 and the other being its complement.
These subsamples contain about the same number of measurements, therefore our ability
to identify the periodic variation induced by the potential second planet should be the same in both of them
(the sampling is roughly similar). 
The normalized power at the peak in the periodogram goes from 3.9 (arbitrary units) for observations
taken before JD=2454500 to 10.1 for observations taken after this date, the power increases for each new added data
point. The period stays constant within one sigma in the two datasets. It is worth underlining that the uncertainty 
on the period is small also for the two separate subsamples: less than $10^{-2}$ days. 

\begin{figure}
  \centering
  \includegraphics[width=9cm]{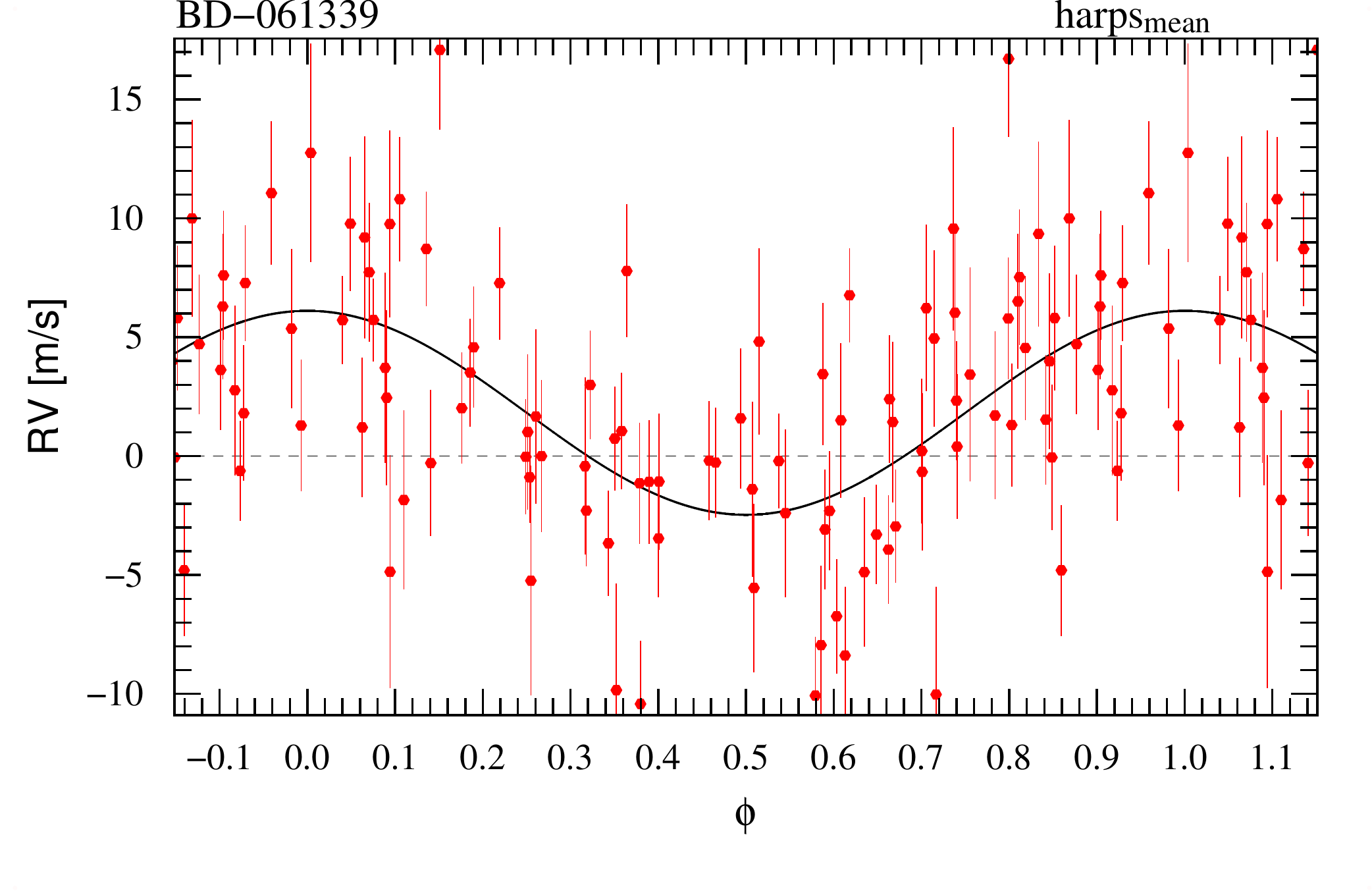}
     \caption{Radial velocity as function of the orbital phase of the short-period planet of the star BD-061339.}
     \label{bd061339_pl2}
 \end{figure}

Adding the fact that we see no correlation between the RV residuals to the one-planet fit and the FWHM of the CCF or the bisector span,
nor do we see significant power in the periodogram of these parameters around the $\approx 3.9$ days period,
we gain confidence that the second signal, which is becoming more evident as the star becomes less active,
is indeed a planet. The RV as a function of the orbital phase is shown in figure \ref{bd061339_pl2}.

This case is nice and instructive, and clearly shows how a planetary signal emerges from the noise of
an active star when the activity level drops. It also reminds us that "active" stars are not necessarily always
active, and that if at some time they are inappropriate for a planet search because of their
high level of activity, at other times detection of planets around these stars may become possible.



\begin {center}
\begin{table*}
\caption{Orbital parameters for the planets. 
$P$: period, $T$: time at periastron passage, $e$: eccentricity, $\omega$: argument of periastron, $K$: semi-amplitude of the radial velocity curve, 
$m \sin{i}$ : projected planetary mass, $a$: semi-major orbital axis, $V$: average radial velocity, $N_{\mathrm{meas}}$: number of measurements, 
$<\delta$\,RV$>$: average uncertainty on the radial velocity measurements, $\sigma$ (O-C) : root mean square of the residuals from the Keplerian fit. 
}
\begin{minipage}[t]{\columnwidth}
\label{orbittab}
\centering
\renewcommand{\footnoterule}{}
\begin{tabular}{l l r r r r r r r r r r}
\hline\hline
\multicolumn{2}{l}{\bf Parameter} &\hspace*{-2mm} & \bf HD\,103774b  & \hspace*{-1mm} & \bf HD\,109271b & \hspace*{-1mm} & \bf HD\,109271c & \hspace*{-1mm} & \bf BD$-061339$b & \hspace*{-1mm} & \bf BD$-061339$c\\
\hline

$P$ & [days] & 							& 5.8881 $\pm$ 0.0005 & 	&7.8543 $\pm$ 0.0009 & 	&  30.93 $\pm$ 0.02 & 	& 3.8728 $\pm$ 0.0004 	&	& 125.94 $\pm$ 0.44	\\
$T$ & [JD-2400000] & 					& 55675.4 $\pm$ 2.0 & 		& 55719.1 $\pm$ 3.6 & 	& 55733 $\pm$8 & 		& 55220.50 $\pm$ 0.10 	&	& 55265.2 $\pm$ 17.6	\\
$e$ & & 								& 0.09  $\pm$ 0.04 & 		& 0.25 $\pm$ 0.08 & 	& 0.15 $\pm$0.09& 		& 0.0 (fixed) 			&	& 0.31 $\pm$ 0.11		\\
$\omega$ & [deg] & 						& -42$\pm$64 & 			& -64$\pm$20 & 		& 4$\pm$ 120 & 		& 0.0 (fixed) 			&	& 41 $\pm$ 55			\\ 
$K$ & [m s$^{-1}$] & 					& 34.3 $\pm$ 1.8 & 			& 5.6 $\pm$ 0.5 & 		& 4.9 $\pm$ 0.4 & 		& 4.4 $\pm$ 0.6		&	&  9.1 $\pm$2.9		\\
$m \sin{i}$ & [$M_{\mathrm{Jup}}$] & 		& 0.367 $\pm$ 0.022 & 		&  0.054 $\pm$ 0.004& 	& 0.076 $\pm$ 0.007 & 	& 0.027 $\pm$ 0.004	&	& 0.17 $\pm$ 0.03		\\ 
$m \sin{i}$ & [$M_{\mathrm{Earth}}$] & 		& 116 $\pm$ 7 & 			& 17 $\pm$ 1 & 		& 24 $\pm$ 2 & 		& 8.5 $\pm$ 1.3		&	& 53 $\pm$ 8			\\ 
$a$ & [AU] & 							& 0.070 $\pm$ 0.001 & 		& 0.079 $\pm$ 0.001 & 	& 0.196 $\pm$ 0.003 & 	& 0.0428 $\pm$ 0.0007	&	& 0.435$\pm$0.007		\\
\hline
$V$ & [km s$^{-1}$] & 					& -3.049$\pm$0.003 & 		& -4.903$\pm$0.001 & 	& -4.903$\pm$ 0.001 & 	& 23.625 $\pm$ 0.003 	&	& 23.625 $\pm$ 0.003	\\
$N_{\mathrm{meas}}$ & &  				& 102 & 					& 87 & 				& 87 & 				& 102 				&	& 102				\\
{\it Time span}& [days] &    				& 2734 & 					& 2683 & 				& 2683 & 				& 2955 				&	& 2955				\\

$<\delta$\,RV$>$ & [ms$^{-1}$] & 			& 2.6& 					& 0.9 & 				& 0.9 & 				& 2.7 				&	& 2.7					\\ 
$\sigma$ (O-C) & [ms$^{-1}$]    & 			& 11.43 & 					& 2.05 & 				& 2.05 & 				& 4.3 				&	& 4.3					\\
$\chi^2_{\rm red}$ &  & 					& 15.7 $\pm$ 0.6 & 			& 4.6 $\pm$ 0.3 & 		& 4.6 $\pm$ 0.3 & 		& 2.47 $\pm$ 0.23 		&	& 2.47 $\pm$ 0.23		\\

\hline
\end{tabular}

\end{minipage}
\end{table*}
\end {center}

\section{Discussion}

We have identified five planetary signals with $m \sin(i)$ going from 0.6 to 7 times the mass of Neptune.
Assuming they lie in the same region of the radius-mass space as the confirmed transiting planets
(\cite{schneider}), we can estimate their radius and therefore their density. 
With this assumption, we obtain a maximum density for these newly discovered planets that is comparable 
to Neptune's or Jupiter's mean density, making them most likely gaseous planets.

\begin{figure}
  \centering
  \includegraphics[width=9cm]{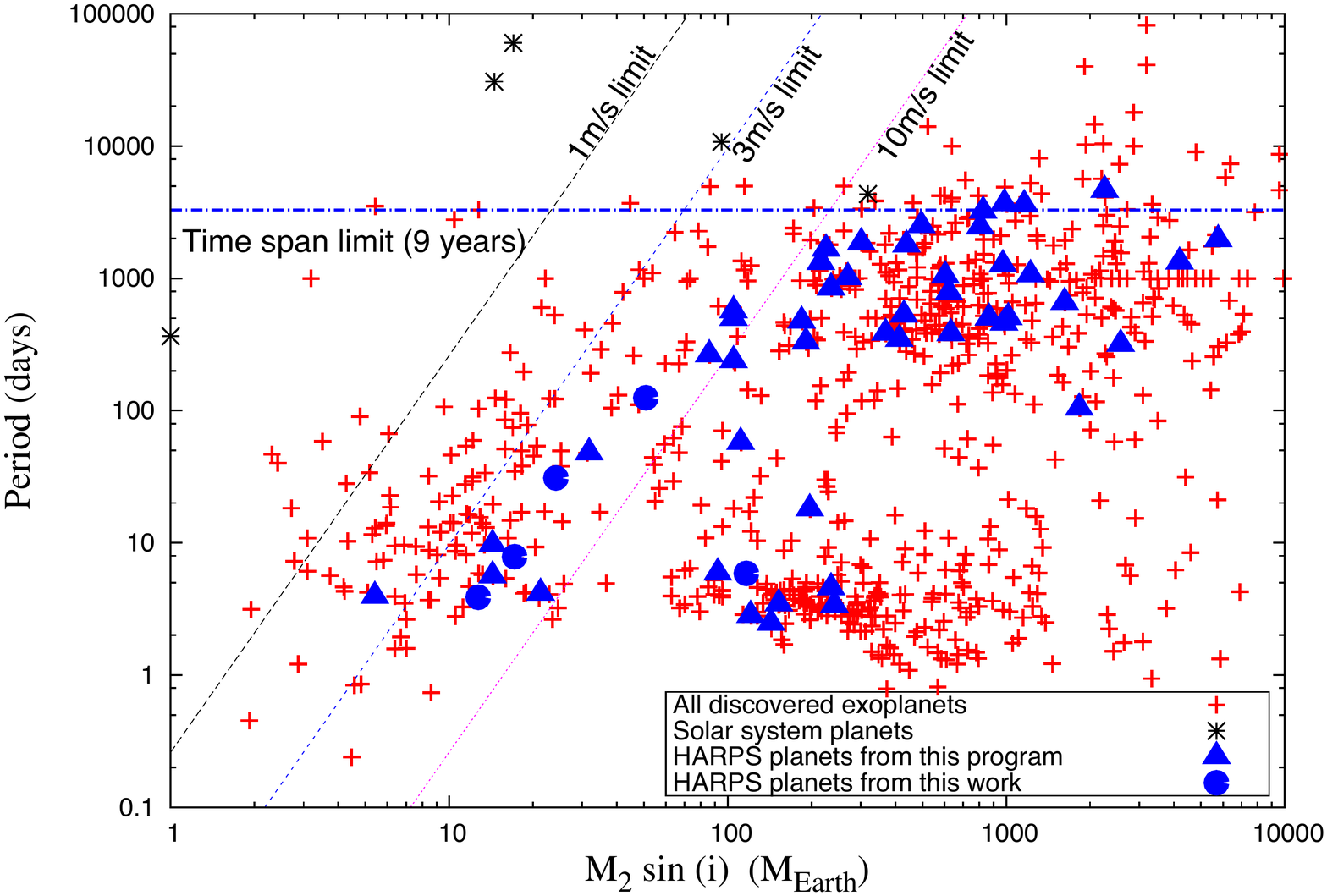}
     \caption{Mass-period distribution of the RV-detected planets. The crosses represent all detected planets, 
     the filled triangles the planets from this program, the asterisks the solar system planets, and the filled circles the planets 
     presented in this work. }
     \label{pm}
 \end{figure}

Observing the current detections in the mass-period diagram (figure \ref{pm}), we see that the planets presented in this paper 
are close to the detection limit of our survey, of about $\approx 3ms^{-1}$, and that two of them fall in the 
low-populated region of intermediate periods, between $\approx 10$ and $\approx 200$ days.
This region is indeed mostly occupied by planets detected in multiple planetary systems.
It makes us wonder whether the presence of more than one planet breaks a certain symmetry that then allows
the planet to occupy a range of otherwise "forbidden" orbital periods.

%
The periods of the two Neptune-mass planets orbiting the star HD109271 are near a 1:4 resonance, 
but this would require a thorough dynamical analysis to be confirmed.
The planetary system orbiting the star HD109271 is very similar to the HD69830 system (\cite{lovis06}):
the periods of the planets $b$ and $c$ differ only by 10$\%$ and 2$\%$ respectively, 
while the planet masses differ by less than a factor of two, which is an indication that there seem to be 
some "preferred architectures" among planetary systems. Whether this is real or an effect of our
limited understanding of the reality is an open question.\\

%
The outer planet of the system orbiting the star BD-061339 is particularly interesting because it falls right in the middle of the habitable zone (HZ) of its star.
Following the method outlined by Kane (2012), we estimate the habitable zone of the star to extend from 0.5 to 0.9 AU.
The orbit of planet $c$, even considering its relatively high eccentricity of 0.31, lies entirely within the HZ.
Depending on the albedo (in the range 0.4 - 0.9) and on the atmospheric model (efficient / inefficient heat transport) the surface temperature of the planet can be between 0 $^oC$ and 100 $^oC$. 
Although suggestive, this is a simplistic approach however, because it does not take into account that the albedo as well as the stellar flux vary with wavelength.
Moreover, due to its mass, the planet is likely to be a gas planet, making conditions for life probably more difficult.


\section{Conclusions}

We have presented new results from our nine-year low-precision ($2-3ms^{-1}$) survey of a volume-limited sample
of the solar neighborhood with HARPS. We detected five exo-planets orbiting three low-activity, slowly rotating solar-type stars.
Four of the detected planets are in two multi-planet systems. The planets have $m \sin(i)$ ranging from 0.6 to 7 Neptune masses,
the lightest being in the super-Earth mass range, 
underlining that our survey, albeit limited in precision because of a strategical choice, has the capability
to detect Neptune-mass and in some cases even super-Earth planets. 

The star BD-061339 hosts two planets: in the case that $\sin(i)=1$, they would be
a super-Earth, possibly in a tidally locked configuration due to its proximity to the host star, and
a planet of $\approx$ three Neptune masses in the habitable zone. The case presented by this star is particularly instructive because it shows 
how one of the signals became more clear with time, as the stellar activity was diminishing.

The star HD103774, hosting one planet with $m \sin(i)$ roughly the mass of Neptune, 
has also shown signs of varying activity during the nine years 
of our monitoring campaign. A reminder that the stellar activity changes with time, and stars that at one time
are unsuitable for high-precision RV measurements, might become more quiet at another time 
and offer the possibility to acquire precise RV data.

The star HD109271 hosts two planets with $m \sin(i)$ approximately the mass of Neptune and with orbital periods and $m \sin(i)$ very similar to  planets $b$ and $c$ around the star HD698430  (\cite{lovis06}). This fact suggests that some planetary architectures might be 
"preferred" in Nature.

Two out of the five planets we detected have periods in the  range between 10 and 200 days, 
a region where planet occurrence is generally low. Interestingly, most of the
planets with periods in this range belong to multiple planetary systems.

Owing to the small amplitude of the signals and to the proximity of their amplitudes, 
approximately 100 data points per system were necessary 
to achieve the current detections, which supports ample time coverage and
high-cadence monitoring as the key elements to access some of the most interesting planetary detections.

\begin{acknowledgements}

We thank the staff of the La Silla Observatory at ESO in Chile for their passionate and professional support.\\
This research has made use of the VizieR catalogue access tool, CDS, Strasbourg, France. The original description of the VizieR service was published in A\&AS 143, 23.

NCS acknowledges the support by the European Research Council/European Community under the FP7 through Starting Grant agreement number 239953, as well as from Funda\c{c}\~ao para a Ci\^encia e a Tecnologia (FCT) through program Ci\^encia\,2007 funded by FCT/MCTES (Portugal) and POPH/FSE (EC), and in the form of grants reference PTDC/CTE-AST/098528/2008 and PTDC/CTE-AST/098604/2008.

We thank the referee Martin K\"urster for the helpful comments which contributed to improve this paper.

\end{acknowledgements}


\end{document}